\documentclass[12pt]{article}
\usepackage{epsfig}
\def\be{\begin{equation}}
\def\ee{\end{equation}}
\def\bea{\begin{eqnarray}}
\def\eea{\end{eqnarray}}

\def\gam5{\gamma_5}

\def\sigmaprim2{\sigma^{\prime 2}}
\def\piprim2{\vec{\pi}^{\prime 2}}

\def\mN{m_{\mbox{\tiny N}}}
\def\eV{e_{\mbox{\tiny V}}}
\def\eA{e_{\mbox {\tiny A}}}
\def\gA{g_{\mbox{\tiny A}}}
\def\tildegA{\tilde{g}_{\mbox{\tiny A}}}
\def\muV{\mu_{\mbox{\tiny V}}}

\newcommand{\boldp}{\mbox{\boldmath $p$}}
\newcommand{\boldk}{\mbox{\boldmath $k$}}

\begin{document} 
\begin{titlepage} 


\begin{center}
\vspace{5mm}
\Large{\bf Radiative corrections to anti-neutrino proton scattering at low energies }
\\
\vskip 0.5cm  
{\large U. Raha, F. Myhrer and K.Kubodera  }
\vskip 0.5cm 
{\large Department of Physics and Astronomy,
University of South Carolina, 
Columbia, SC 29208 }
\end{center}
\vskip 1cm
\begin{abstract}

For the low-energy anti-neutrino reaction,
$\bar{\nu}_e + p \to e^+ + n$, 
which is of great current interest in connection 
with on-going high-precision neutrino-oscillation experiments,
we calculate the differential cross section 
in a model-independent effective field theory (EFT),
taking into account radiative corrections of order $\alpha$. 
In EFT,  the short-distance radiative corrections are 
subsumed into well-defined low-energy constants
the values of  which can in principle
be determined from the available neutron beta-decay data. 
In our low-energy EFT, the order-$\alpha$ radiative corrections 
are considered to be of the same order as the nucleon recoil corrections, 
which include the ``weak magnetism" contribution. 
These recoil corrections have been evaluated as well. 
%
We emphasize that EFT allows for a systematic evaluation of 
higher order corrections, providing estimates of theoretical uncertainties
in our results.  

\end{abstract}
\end{titlepage}

\newpage 

\section{Introduction}
Low-energy anti-neutrinos from nuclear reactors are well suited 
to determine the neutrino mixing angle $\theta_{13}$,
which is important for the search of CP violation in the leptonic sector; 
see, e.g., Refs.~\cite{White2003,Minakata03}. 
The Double-Chooz~\cite{chooz2003},  
Daya Bay~\cite{dayabay}, and RENO Collaborations~\cite{RENO} 
are aiming to measure $\theta_{13}$
with very high precision
with the use of  $\bar{\nu}_e$'s produced in nuclear reactors.
The present upper bound 
to this quantity reported by the Chooz~\cite{chooz2003} and
MINOS~\cite{minos2010}  Collaborations
is: $\theta_{13} < 11.4^\circ $.

The Double-Chooz, Daya Bay and RENO experiments
monitor the inverse beta-decay reaction on a hydrogen target
\be
\bar{\nu}_e + p \to e^+ + n \label{DCDB}
\label{eq:reaction}
\ee
for a known anti-neutrino energy flux.
The positron yield is measured as a function of the positron energy.
An accurate extraction of the mixing angle $\theta_{13}$
from an analysis of the measured positron yield
requires a precise knowledge of the radiative corrections (RCs).
In earlier papers~\cite{Vogel84,fayans85,FK04}, 
the relevant RCs were evaluated in the theoretical framework 
developed by Sirlin and Marciano~\cite{Sirlin1974,MS86}.
In this framework, to be referred to as the S-M approach,
the RCs of order $\alpha$ are decomposed into so-called outer 
and inner corrections.
The outer correction is a universal function of the lepton energy
and is independent of the details of hadron physics,
whereas the inner correction is influenced by short-distance physics
and the hadron structure.
The inner corrections coming from $\gamma$ and weak-boson loop
diagrams are divided into high-momentum and low-momentum parts.
The former is evaluated in the current-quark picture, 
while the latter is computed with the use of
the phenomenological electroweak-interaction form factors of the nucleon. 
Although the estimates of inner corrections in the S-M formalism
are considered to be reliable to the level of accuracy quoted in the literature,
the possibility that these estimates may involve some degree
of model dependence is not totally excluded. 

We present here a calculation of the RCs to order $\alpha$
based on effective field theory (EFT).
We use heavy-baryon chiral perturbation theory (HB$\chi$PT),
which is an effective low-energy theory of QCD, see e.g. Ref~\cite{bkm95}.
In HB$\chi$PT  the short distance hadronic and electroweak processes  
are subsumed into a well-defined set of  {\it low-energy constants} (LECs). 
In other words, these LECs
systematically
parameterize the inner corrections of the S-M approach. 
Therefore, insofar as there are enough sources of information
to determine the values of these LECs, 
HB$\chi$PT leads to model-independent results 
with systematic estimates of higher-order corrections.
The use of HB$\chi$PT to calculate electroweak transition amplitudes
for  the nucleon and few-nucleon systems were pioneered 
in  Refs.~\cite{Rho,PMRa, PMRb}, and subsequently 
there have been many important developments.
In Ref.~\cite{Ando2004},
we presented the first ever EFT-based calculation
of RCs for the neutron $\beta$-decay process, $n\to p+e^-+\!\bar{\nu}_e$. 
Because in HB$\chi$PT the nucleons are treated as point-like,
it is expected on general grounds that the 
order-$\alpha$ RCs are common between 
neutron $\beta$-decay and inverse $\beta$-decay.  
Meanwhile, it should be mentioned 
that, in the counting scheme
adopted here and in Ref.\cite{Ando2004},
the order-$\alpha$ RCs are of the same order as 
the $\mN^{-1}$ nucleon-recoil corrections
including the ``weak magnetism" contributions,\footnote{
The importance of the nucleon-recoil corrections was 
emphasized by, e.g., Vogel and Beacom~\cite{VB1999}.}
and hence a consistent  
EFT calculation
should include these recoil corrections simultaneously.
We present here such an EFT calculation,
taking advantage of the fact that the 
$\mN^{-1}$-expansion 
is a natural part of our counting scheme 
and thus 
dictates how to incorporate recoil corrections order by order (see later in the text). 

Since exactly the same LECs are involved in 
the EFT calculations of inverse $\beta$-decay and neutron $\beta$-decay,
we can  in principle use the existing neutron $\beta$-decay data
to determine those LECs
and make a model-independent estimation of RCs 
for the inverse $\beta$-decay,
\underline{provided that} the $\mN^{-1}$ recoil corrections 
are properly taken into account.
In this connection,
we note that an attempt has been made
in the literature~\cite{Vogel84,VB1999}
to directly relate the neutron decay rate 
with the inverse $\beta$-decay cross section,
assuming that the order-$\alpha$ corrections
(RCs and recoil corrections combined)
are common between these processes.
As mentioned, this assumption is justified as far as the genuine
RCs of order $\alpha$ is concerned.
However, as described later in the text,
our calculation shows differences 
between the $\mN^{-1}$ corrections for inverse $\beta$-decay, 
Eq.~(\ref{eq:reaction}),
and those for neutron $\beta$-decay. 
We therefore caution against writing the cross section for the 
reaction in Eq.~(\ref{eq:reaction}) in terms of the neutron mean life,
$\tau_n$, as advocated in Refs.~\cite{Vogel84,VB1999}. 
  
This paper is organized as follows.
In section 2 we explain a theoretical framework to be used and 
present the results for the order-$\alpha$ RCs.
In section 3 we consider the recoil corrections and compare our
results with an earlier work\cite{VB1999}. 
Section 4 gives a summary of our calculations and conclusions. 
The appendix describes some technical details concerning 
the HB$\chi$PT treatment of the infrared singularity.

\section{The QED corrections}

We use here  essentially the same theoretical framework as
in Ref.~\cite{Ando2004}, in which
we calculated RCs for neutron $\beta$-decay.
We therefore give only a brief recapitulation of our formalism,
relegating details to Ref.~\cite{Ando2004}.

Our calculation is based on the $\bar{Q}/\Lambda_{\chi}$-expansion scheme, 
where $\bar{Q}$ $\sim$ $E_{\nu}\!-\!\Delta_N\!-\!m_e$ 
($\Delta_N=m_n-m_p$) 
represents a typical 
four-momentum transfer for  
incident low-energy reactor anti-neutrinos ($E_{\nu}\leq 10$ MeV), and 
$\Lambda_{\chi}\simeq 4\pi f_{\pi}\approx1$ GeV ($f_{\pi}=92.4$ MeV is the
pion-decay constant) is  the chiral scale. 
It is to be noted that the expansion parameter in our scheme 
is very small and that, 
as explained in more detail below,
the lowest order recoil corrections $\sim \bar{Q}/\mN$ are 
of the same order as the lowest order radiative corrections; 
{\it viz.}, 
$\bar{Q}/\mN\sim\alpha/(2\pi)\sim\bar{Q}/\Lambda_{\chi}\sim10^{-3}$, 
where $\mN=(m_p\!+\!m_n)/2$. 


The {\it leading order} (LO) transition matrix element for the inverse $\beta$-decay,
Eq.(\ref{DCDB}), is evaluated ignoring nucleon recoil and radiative corrections. 
The {\it next-to-leading order} (NLO) corrections in our counting scheme
are  the recoil corrections ($\sim Q/\mN$) and  
the  radiative corrections ($\sim \alpha/(2\pi)$).
The recoil corrections, which include the ``weak magnetism" term, 
will be specified in Eq.(\ref{eq:Lnnee}) below. 
For the sake of the transparency of presentation, 
we shall in this paper separate these corrections from the 
$\mN^{-1}$ (kinematic) corrections to the phase-space. 

The effective lagrangian relevant to our calculation includes 
the relativistic leptonic weak interaction current 
and the LO and NLO heavy-baryon lagrangian 
\begin{eqnarray}
{\cal L}_{eff} &=& {\cal L}_{QED} + {\cal L}_{NN} +{\cal L}_{NN\psi\psi } \; , 
\end{eqnarray}
where  
\begin{eqnarray}
{\cal L}_{QED}\!\!&=&\! -\frac{1}{4} F^{\mu \nu}\! F_{\mu \nu} 
-\frac{1}{2\xi_A}(\partial\!\cdot\! A)^2 + 
\left(\! 1\!+\!\frac{\alpha}{4\pi}e_1\!\right)\! \bar{\psi}_e (i\gamma\!\cdot\! D) \psi_e 
+ m_e \bar{\psi}_e  \psi_e +
\bar{\psi}_\nu i\gamma\!\cdot\!\partial\psi_\nu  , 
\label{eq:Lqed}
\\ 
{\cal L}_{NN}\!\!&=&\!\bar{N} 
\left[1+\frac{\alpha}{8\pi} e_2 (1+\tau_3) \right] 
(iv\cdot D) N \; , 
\label{eq:Lnn}\\ 
{\cal L}_{NN\psi\psi } \!&=&\!\!\!\! -\!\left(\frac{G_F V_{ud}}{\sqrt{2}}\right) \!
\left[ \bar{\psi}_e \gamma_\mu (1-\gamma_5)\psi_\nu \right] \!
\left\{ \bar{N}  \tau^+ \!
\left[ \left( 1\!+\!\frac{\alpha}{4\pi} \eV \right) 
 v^\mu - 2g_A\! \left( 1\!+\!\frac{\alpha}{4\pi} \eA \right)  S^\mu \right]\! N 
\right.
\nonumber \\
\!&+ &\!\!\!
\left. \!\!
\frac{1}{2m_N} \bar{N} \tau^+\!\!
\left[ i(v^\mu v^\nu\!\! -\! g^{\mu \nu} )(\stackrel{\leftarrow}{\partial} 
\!-\! \stackrel{\rightarrow}{\partial} )_\nu 
\!-\!2i\muV\!\left[ S^\mu\! , S\!\cdot\! (\stackrel{\leftarrow}{\partial} 
\!+\!\stackrel{\rightarrow}{\partial} ) \right] 
\!-\!2i\gA v^\mu S\!\cdot\! (\stackrel{\leftarrow}{\partial} 
\!-\!\stackrel{\rightarrow}{\partial} ) \right]\! N \right\} .
\nonumber \\ &&
\label{eq:Lnnee}
\end{eqnarray}
${\cal L}_{QED}$ in Eq.(\ref{eq:Lqed})
is the usual QED lagrangian, where    
$F_{\mu \nu} \!=\! \partial_\mu A_\nu \!-\! \partial_\nu A_\mu$,  
and $D_\mu= \partial_\mu \!+\! ieA_\mu$ 
is the covariant derivative;
for the gauge parameter $\xi_A$,
we use here $\xi_A =1$ (Feynman gauge). 
${\cal L}_{NN}$ is the heavy-nucleon lagrangian including the  
photon-nucleon interaction, and 
${\cal L}_{NN\psi\psi }$ is the low-energy LO and NLO 
current-current weak interaction. 
We give in Eq.(\ref{eq:Lnnee}) the explicit forms 
of NLO nucleon-recoil terms dictated by HB$\chi$PT. 
In the above,  $g_A = 1.267$ is the axial coupling constant, 
while $v_\mu$ is the nucleon velocity vector, and 
$S^\mu$ is the nucleon spin; they satisfy $v\cdot S =0$.
We choose here $v^\mu = (1, \vec{0} )$ and
$S^\mu = (0, \vec{\sigma} /2)$. 
In the NLO part of the lagrangian the nucleon isovector magnetic moment 
$\muV = \mu_p\!-\!\mu_n = 4.706$. 
The low-energy constants (LECs), $e_1$, $e_2$, $\eV$ and $\eA$, 
are counter-terms which regulate 
the ultraviolet (UV) divergences of the virtual photon-loop diagrams.  
These LECs incorporate the short-range radiative physics 
that is not probed in a low-energy process.
The LECs, $e_1$ and $e_2$, are 
related to the wave-function renormalization factors of the 
positron and proton, respectively. 
The LECs, $\eV$ and $\eA$, are related 
to the Fermi and Gamow-Teller amplitudes. 
%
The Fermi coupling constant,
$G_F = 1.166 \times 10^{-5}$ GeV$^{-2}$, is determined from muon decay,  
and the CKM matrix element, $|V_{ud}| = 0.97418 \pm 0.00027$, is 
given by the PDG~\cite{PDG2008}.

For later convenience, we rearrange the LECs 
in Eq.(\ref{eq:Lnnee}) by rewriting 
the hadronic part in the first line in Eq.(\ref{eq:Lnnee}) as 
\begin{eqnarray} 
\bar{N}\tau^+ \!\left[ v^\mu \!-\!2\tildegA S^\mu \right] \!N 
+\left(\frac{\alpha}{4\pi}\right)\!\eV
\bar{N}\tau^+ \!\left[ v^\mu \!-\!2\tildegA S^\mu \right]\! N 
+ {\cal O}(\alpha^2) \; ,
\nonumber
\end{eqnarray}
where we have introduced the redefined axial coupling constant, 
$\tildegA = \gA[1\!+\!\frac{\alpha}{4\pi}(\eA\!-\!\eV)]$. 
As in the neutron $\beta$-decay case~\cite{Ando2004}, 
to the order of our concern,
$\gA$ can always be replaced by $\tildegA$. 
This also applies to the NLO recoil contributions since the $\mN^{-1}$ 
recoil corrections are of the same order as 
order-$\alpha$ corrections in the adopted counting scheme. 
The order-$\alpha$ radiative corrections to the nucleon 
magnetic moments are for the same reason of higher order in our scheme 
and hence neglected in this work. 

In this paper we derive a model-independent expression 
for the lowest order radiative and recoil corrections to the reaction,
$\bar{\nu}_e(p_\nu)+p(p_p) \to e^+(p_e) + n(p_n)$,
where the four-momentum of each particle is indicated
in the parentheses.
We shall concentrate on an experimental setup
in which none of the particle spins are monitored by the detector. There is
one subtle aspect of the above reaction which deserves some discussion. In
experiments, the final state positron will always be accompanied by (often 
undetected) soft bremsstrahlung photons. 
If the bremsstrahlung photon energy, 
$E_\gamma$ is less than the detector resolution, $\Delta$, the
energy recorded by the detector is the sum of the
actual outgoing positron energy, $E_e$, 
and the bremsstrahlung photon energy, $E_\gamma$;
i.e., $E=E_e+E_\gamma$ is what is measured as the
``detected positron'' energy, with the corresponding 
``detected positron'' momentum being $| \boldp |=\sqrt{E^2-m^2_e}$. 
The two processes we evaluate are $\bar{\nu}_e+p \to e^+ + n$ and
$\bar{\nu}_e+p \to e^+ + \gamma + n$.
Due to the finite detector resolution the second bremsstrahlung process 
is not observed; it only contributes to the RCs of the first process, 
i.e., the soft 
bremsstrahlung photons 
are an integral part of the ``detected positron". 
Thus, the first process has become, 
$\bar{\nu}_e(p_\nu)+p(p_p) \to e^+(p) + n(p_n)$, where 
the positron momentum $p_e$ has been replaced by $p$ 
in order to indicate that the soft bremsstrahlung process has been incorporated 
into this ``effective" reaction.  
The cross section for this ``effective"  
reaction is given in terms of the effective 
invariant amplitude ${\cal M}$:
\begin{eqnarray}
{\rm d}\sigma  \!\!\!&=& \!\!\!
\frac{ 1 }{4m_pE_\nu}  \int \frac{{\rm d}^3\boldp}{(2\pi)^32E}   
\frac{{\rm d}^3\boldp_n}{(2\pi)^32E_n}(2\pi)^4 \delta^{(4)}
(p_\nu\!+\!p_p\!-\!p\!-\!p_n)
\,\frac{1}{2} \sum_{\rm spins} |{\cal M}|^2 
\nonumber \\ &=& \!\!
\left( \frac{G_F V_{ud} }{\sqrt{2} } \right)^2  \!\!
f(E) 
\left[(1\!+\!3g_A^2)\,{\cal G}_1(\beta) + (1\!-\!g_A^2)\, 
{\cal G}_2(\beta) \beta \cos\theta_e \right] 
{\rm d}(\cos\theta_e) \, , 
\label{eq:NLO}
\end{eqnarray}
where $\beta = | \boldp |/E = \sqrt{E^2-m^2_e}/E$ is the velocity 
of the outgoing ``detected positron'' for a given incident (anti-)neutrino beam 
energy, $E_\nu$ and detector reading, $E$;  
 $\cos (\theta_e) = \hat{\boldp}_\nu\!\cdot\hat{\boldp}$, and 
$f(E)$ is the phase-space factor to be discussed
later in the text (see, Eq.~(\ref{eq:PS-recoil})).  
The two velocity-dependent functions, ${\cal G}_i(\beta)$ ($i = 1, \ 2$), 
are written up to NLO as
\begin{eqnarray}
{\cal G}_i(\beta) \!&=&\!\! 1+\frac{\alpha}{2\pi}\,{\cal G}_i^{rad}(\beta) 
+\frac{1}{\mN}\,{\cal G}_i^{recoil}(\beta) \; , 
\label{eq:Gi}
\end{eqnarray}
Here ${\cal G}_i^{rad}(\beta)$ (see, Eqs.~(\ref{eq:G1}) and (\ref{eq:G2})) 
represent the lowest-order 
radiative corrections, and ${\cal G}_i^{recoil}(\beta)$ 
(see, Eq.~(\ref{eq:L-recoil})), 
which will be evaluated in the next section,
represent the recoil corrections 
arising from  the lagrangian in Eq.(\ref{eq:Lnnee}). 
The calculation of the function ${\cal G}_i^{rad}(\beta)$  is described next.
%

\begin{figure}[t]
\begin{center}
\epsfig{file=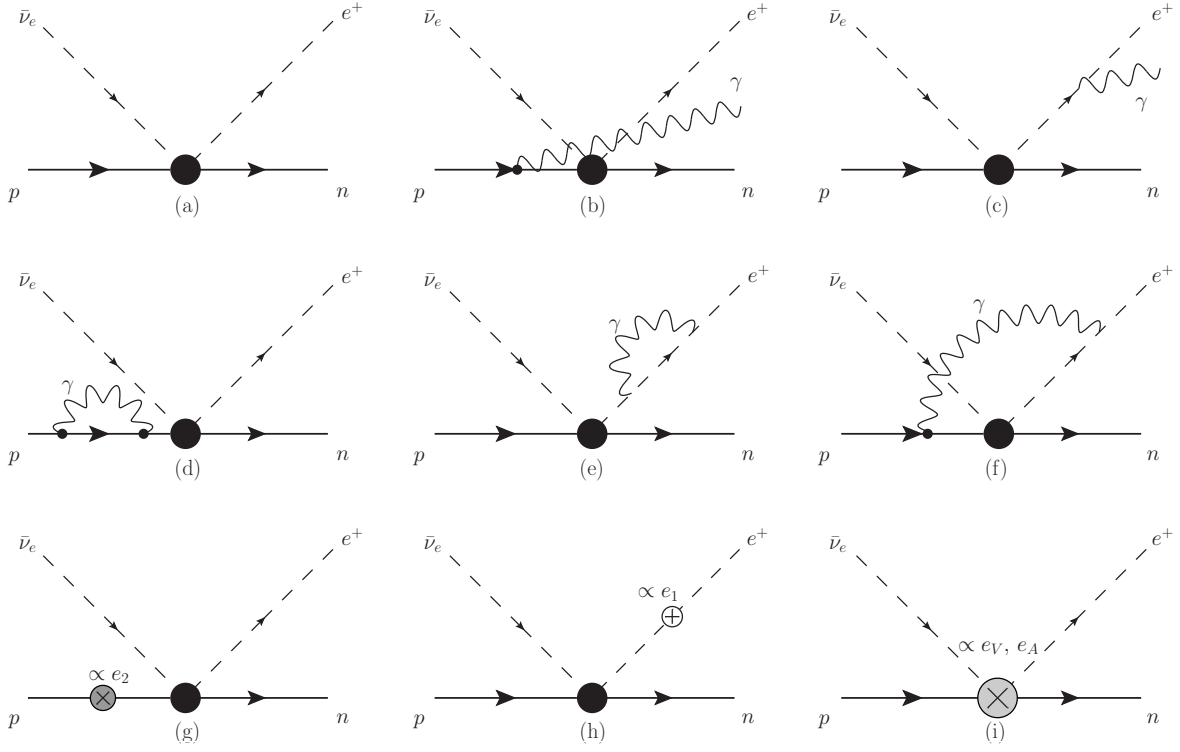,width=16cm}
\end{center}
\caption{Feynman diagrams contributing to the cross section which
           includes ${\mathcal O}(\alpha)$  QED corrections:
           (a) ---  leading order (LO) Born amplitude; 
           (b) and (c) ---    bremsstrahlung amplitudes;
            (d), (e) and (f) ---  virtual photon-loop diagrams;
           (g) and (h) --- counter-term amplitudes involving the LECs
           $e_1$ and $e_2$, respectively;
           (i) ---  counter-term amplitudes involving both of the LECs, $\eV$ and $\eA$. 
           The $\mN^{-1}$ correction is represented by diagram (a) but 
           with the use of the vertex that arises from the $\mN^{-1}$ (NLO) part 
           of the lagrangian in Eq.(\ref{eq:Lnnee}). 
            To the order of our concern we do not consider diagrams 
            in which a photon couples to the nucleon magnetic moments
             since this is ${\cal O}(\alpha /m_N)$ or NNLO. }
\label{fig:feynmanQED}
\end{figure}

For the analysis of the radiative corrections, we explicitly distinguish
between the outgoing positron and the bremsstrahlung photon.
There are two distinct categories of radiative corrections,  
the bremsstrahlung  
and the virtual photon loop corrections. 
The corresponding Feynman diagrams are shown in Fig.~\ref{fig:feynmanQED}. 
%
Since ${\cal O}(\alpha)$ and 
${\cal O}(m_N^{-1})$ are of the same order in our counting scheme, 
the invariant matrix element, ${\cal M}_{br}$ for 
bremsstrahlung is evaluated assuming $E_n=m_n$. 
The differential cross section for the radiative process,
$\bar{\nu}_e(p_\nu)\!+\!p(p_p) \!\to\!
e^+(\tilde{p}_e)\!+\!n(p_n)\!+\!\gamma(\tilde{k})$\footnote{ 
The four-momenta,
  $\tilde{p}_e$ and $\tilde{k}$ denote momenta in the static nucleon 
approximation, i.e., 
$p_e=\tilde{p}_e-\mathcal{O}(m^{-1}_N)$, 
$k=\tilde{k}-\mathcal{O}(m^{-1}_N)$,
  and correspondingly, $E =\tilde{E}-\mathcal{O}(m^{-1}_N)$.}, 
is given by
\begin{eqnarray}
{\rm d}\sigma_{br}(\bar{\nu}_e p \to e^+ n \gamma ) &=&
\frac{1}{8m_pm_nE_\nu}\int \frac{{\rm d}^3\tilde{\boldp}_e}{(2\pi)^32\tilde{E}_e}
\frac{{\rm d}^3\tilde{\boldk}}{(2\pi)^32\tilde{E}_\gamma} 
\nonumber \\ &\times & (2\pi)\delta(E_\nu -\Delta_N -\tilde{E}_\gamma-\tilde{E}_e)
\frac{1}{2}\sum_{spins}|{\cal M}_{br} |^2 \; , 
\label{eq:sigmabr}
\end{eqnarray}
where $\tilde{E}=\tilde{E}_e+\tilde{E}_\gamma=E_\nu-\Delta_N$ is the 
maximum energy of the ``detected positron'' in 
the static 
nucleon approximation, i.e., $m_e\leq E\leq \tilde{E}$. The bremsstrahlung 
matrix element squared with the static neutron is 
\begin{eqnarray}
\frac{1}{2} \sum |{\cal M}_{br}|^2 &=& 
\left(\frac{eG_FV_{ud}}{\sqrt{2}}\right)^2 
\left( \frac{ 32 m_n  m_p  \tilde{E}_e  E_\nu}
{ \tilde{E}_\gamma (\tilde{p}_e\cdot \tilde{k})}\right) 
\nonumber \\ & \times & 
\!\!\!\!\left\{ - 
\left[ \frac{(1+3g_A^2) (\tilde{p}_e\cdot \tilde{k})}{ \tilde{E}_\gamma }\right] 
\left[ 1+\left(\frac{1-g_A^2}{1+3g_A^2}\right)
\frac{\tilde{\boldp}_e\cdot\boldp_\nu}{\tilde{E}_eE_\nu }\right] 
\right.
\nonumber \\ &+& \!\!\!
(1\!+\!3g_A^2) 
\Big[ \frac{2\tilde{E}_e^2 +\tilde{E}_e \tilde{E}_\gamma+
\tilde{\boldp}_e\cdot\tilde{\boldk}+ \tilde{E}_\gamma^2 }{\tilde{E}_e} 
- \ \frac{ m_e^2 \tilde{E}_\gamma (\tilde{E}_e+\tilde{E}_\gamma )  }
{\tilde{E}_e(\tilde{p}_e\cdot \tilde{k})}\Big]
\nonumber \\ &+&\!\!
(1-g_A^2)\Big[(\tilde{\boldp}_e\cdot\boldp_\nu)\left(
\frac{2\tilde{E}_e+\tilde{E}_\gamma}{\tilde{E}_e E_\nu} -  
\frac{ m_e^2\tilde{E}_\gamma }{\tilde{E}_e E_\nu (\tilde{p}_e\cdot \tilde{k})}\right)
\Big] 
\nonumber \\ &+&  
\left. 
\!\!(1-g_A^2)\Big[
(\tilde{\boldk}\cdot\boldp_\nu)\left( 
\frac{\tilde{E}_e+ \tilde{E}_\gamma }{\tilde{E}_e E_\nu} 
- \frac{ m_e^2 \tilde{E}_\gamma }{\tilde{E}_e E_\nu (\tilde{p}_e\cdot \tilde{k})}\right)\Big]
\right\} \; . 
\label{eq:Mbrems2}
\end{eqnarray}
%
%
%
We note that the above expression for $\sum |{\cal M}_{br}|^2 $
is identical to that for neutron $\beta$-decay
derived in Ref.~\cite{Ando2004}.
We also remark that Eq.(\ref{eq:Mbrems2})
was derived earlier by Fukugita and Kubota~\cite{FK04},
who used the S-M approach~\cite{Sirlin1974,MS86} and a finite photon mass 
in order to regulate the infrared (IR) singularity.
In the integration over the bremsstrahlung 
photon energy in Eq.(\ref{eq:sigmabr}) the   
maximum photon energy occurs when 
$\tilde{E}_e \!=\! m_e$, i.e.,  
$\tilde{E}^{max}_\gamma = \tilde{E}-m_e$.\footnote{
The approximate integrals considered in Ref.~\cite{Vogel84}
give the analytic expression in Ref.~\cite{FK04} provided
the lower limits of the integrals are changed from 1 MeV to $m_e$.
}
In this context, 
the same question again arises as to 
whether or not the experiment can distinguish 
between the two final states, $n + e^+ + \gamma$ and $n+e^+$.
If the detector resolution in the experimental setup is such that 
one can detect photons with an energy $\tilde{E}_\gamma$ in the interval 
$\Delta \le \tilde{E}_\gamma \le \tilde{E}_\gamma^{max}$, 
we should integrate the bremsstrahlung photon energy 
$\tilde{E}_\gamma = |\tilde{\boldk} |$ from 0 to $\Delta$ 
in Eq.~(\ref{eq:sigmabr}). 
However, if  the experiment is unable to distinguish these two final states,  
we should integrate from 0 to $\tilde{E}_\gamma^{max}$. 
In order to compare with earlier works,
we concentrate here on the latter case. 
%
The integral over the radiative photon spectrum invariably 
gives rise to an IR singularity. 
We use dimensional regularization 
to deal with the IR singularity;  
some details regarding the bremsstrahlung integral are presented in the appendix.
As is well known, the IR singularity appearing in Eq.(\ref{eq:sigmabr}) 
is cancelled by the contributions from virtual photon-loop diagrams 
in accordance with Bloch and Nordsieck~\cite{BN}, see also~\cite{KLN}.
The evaluation of the loop diagrams in 
dimensional regularization can be found in the literature,
see, {\it e.g.,} Ref.~\cite{Ando2004}. 
It is notable that, apart from the so-called ``Coulomb factor" 
$\pi^2/\beta$,
which arises in, e.g. neutron $\beta$-decay from a 
photon-loop diagram, the matrix element given by these virtual photon loops 
is identical to the one in neutron $\beta$-decay. 

The UV-divergencies originating from the photon loop diagrams 
are regulated by the LECs in the lagrangian. 
These LECs are renormalized by the usual effective field theoretical method based on 
dimensional regularization of the loop integrals, see e.g., Refs.~\cite{bkm95,Scherer2003}. 
The finite LECs renormalized at the scale $\mu$ are  
\begin{eqnarray}
e_{V,A}^R(\mu^2)= e_{V,A} -\frac{1}{2}(e_1+e_2) +\frac{3}{2}
\left[ \frac{2}{d-4} -\gamma_E +{\rm ln}(4\pi) +1\right] 
+\frac{3}{2}\,{\rm ln}\left(\frac{\mu^2}{m_N^2}\right) \; .
\nonumber
\end{eqnarray}
The LEC, $\eV^R(\mu^2)$, which was introduced 
by Ando {\it et al.}~\cite{Ando2004} 
in the evaluation of the RC for neutron $\beta$-decay,
subsumes short distance physics not probed at low energies  
and  depends on the regularization scale $\mu$.

%
Combining the bremsstrahlung 
and virtual photon-loops contributions calculated to order $\alpha$, and
noting that
$\tilde{\beta}=\sqrt{\tilde{E}^2-m^2_e}/\tilde{E}=\beta+\mathcal{O}(m^{-1}_N)$,
we obtain, neglecting $\mathcal{O}(m^{-1}_N)$ contributions,  
${\cal G}^{rad}_i(\tilde{\beta})\simeq {\cal G}^{rad}_i(\beta)
$, $i=1$, $2$, 
appearing in Eq.(\ref{eq:Gi}).
Dropping terms of $\mathcal{O}(\alpha\,m^{-1}_N)$, we choose to write the
results in the following form:
\begin{eqnarray}
1+\frac{\alpha}{2\pi}\, {\cal G}_1^{rad}(\beta) &=& 
\left[1\!+\!\frac{\alpha}{2\pi}\,\tilde{e}^R_V (\mu^2)\right]
\left[1\!+\!\frac{\alpha}{2\pi}\delta_{out}(\beta)\right]
\label{eq:G1}
\end{eqnarray}
\begin{eqnarray}
1+\frac{\alpha}{2\pi}\, {\cal G}_2^{rad}(\beta) &=& 
\left[1+\frac{\alpha}{2\pi}\tilde{e}^R_V (\mu^2)\right]
\left[1+\frac{\alpha}{2\pi}\tilde{\delta}_{out}(\beta)\right] \; , 
\label{eq:G2}
\end{eqnarray}
where the ``inner" corrections, 
which are independent of $\beta$, 
are encoded in the LEC $\tilde{e}_V^R(\mu^2)$ and   
defined as 
$\tilde{e}_V^R(\mu^2)= \eV^R(\mu^2) +\frac{5}{4}$. 
The ``outer" radiative corrections constitute the well-known, model-independent, long-distance 
QED corrections that  do not contain any hadronic effects, 
and  are given by 
\begin{eqnarray}
\delta_{out}(\beta) &=& 
3\,{\rm ln}\left(\!\frac{\mN}{m_e}\!\right) + \frac{23}{4} + 
\frac{8}{\beta} \,L\!\!\left(\frac{2\beta}{1\!+\!\beta}\right) 
-\frac{8}{4\beta}\, {\rm ln}^2\!\left(\frac{1\!+\!\beta}{1\!-\!\beta}\right)
\nonumber \\ && 
+ 4 \,{\rm ln}\!\left(\!\frac{4\beta^2}{1\!-\!\beta^2}\!\right) 
\left[ \frac{1}{2\beta} \,{\rm ln}\!\left(\frac{1\!+\!\beta}{1\!-\!\beta}\right)\!-\!1\right] 
+\left(\frac{3\beta}{4}\! +\!\frac{7}{4\beta} \right)
{\rm ln}\!\left(\frac{1\!+\!\beta}{1\!-\!\beta}\right) 
\label{eq:delta-out}
\end{eqnarray}
\begin{eqnarray}
\tilde{\delta}_{out}(\beta) &=& 
3\,{\rm ln}\left(\!\frac{\mN}{m_e}\!\right) +\frac{3}{4} + 
4 \left(\frac{ 1\!-\!\sqrt{1\!-\!\beta^2}}{\beta^2}\right)
+ \frac{8}{\beta} \,L\!\!\left(1\!-\!\sqrt{\frac{1\!-\!\beta}{1\!+\!\beta}} \right) \nonumber\\
&&
+\left( \frac{1}{2\beta}-\frac{3}{8}-\frac{1}{8\beta^2}\right) 
{\rm ln}^2\!\left(\frac{1+\beta}{1-\beta}\right) +\left[ \frac{1}{2\beta}-2 \right]
{\rm ln}\!\left(\frac{1\!+\!\beta}{1\!-\!\beta}\right)  
\nonumber \\ && 
- 4 \left[ \frac{1}{2\beta} \,
{\rm ln}\!\left(\frac{1\!+\!\beta}{1\!-\!\beta}\right) \! -\!1\right] 
{\rm ln}\!\left[\left(\frac{ 1\!+\!\beta }{ 2\beta }\right) 
\frac{\sqrt{1\!+\!\beta}+\!\sqrt{1\!-\!\beta}  }
{ \sqrt{1\!+\!\beta}\!-\!\sqrt{1\!-\!\beta} } \ \right] \; .
\label{eq:deltatilde-out}
\end{eqnarray} 
%
The above expressions for $\delta_{out}$ and $\tilde{\delta}_{out}$
reproduce the results obtained by 
Fukugita and Kubota~\cite{FK04}.
We also note that $\delta_{out} \equiv h(\hat{E}, E_0)$,
where $h(\hat{E}, E_0)$ is the function introduced by Sirlin~\cite{Sirlin2011}.

As mentioned,  $\eV^R(\mu^2)$ also 
appears in the expression for the RC for neutron $\beta$-decay.
Therefore, it is  in principle possible to determine  $\eV^R(\mu^2)$
using relevant high-precision low-energy data involving baryons. 
Due to lack of useful experimental data, 
Ando {\it et al.}~\cite{Ando2004} determined $\eV^R(\mu^2)$
at $\mu=\mN$ by comparing their results for neutron $\beta$-decay 
with those obtained 
in the S-M approach~\cite{Sirlin1974,MS86}.

\vspace{8mm}

\noindent
\section{The  $\mN^{-1}$  recoil corrections}

As mentioned, these corrections have two different origins.
One comes from the lagrangian itself,
and the other arises from the expansion of the kinematic factors in 
the phase-space integral. 
Below we treat these two types of recoil corrections separately and 
compare our results with those in Ref.~\cite{VB1999}. 
It is to be noted again that, 
in evaluating the $\mathcal{O}(m^{-1}_N)$ corrections, 
we can neglect $\mathcal{O}(\alpha)$ radiative effects,
since  $\mathcal{O}(\alpha\, m^{-1}_N)$ terms are of higher order
in our counting scheme. 
One can, therefore, 
assume that the outgoing positron energy, $E_e\approx E$, and correspondingly,
the positron velocity, $\beta_e\approx \beta$.
\vspace{5mm}

\noindent
{\bf Kinematic (phase space) corrections} \\ 

\noindent
The phase space factor, $f(E)$, appearing in Eq.(\ref{eq:NLO}) to the 
lowest order (LO) in the $\mN^{-1}$ expansion is given by
$f(\tilde{E}) = \tilde{E}^2\tilde{\beta}/ \pi$  with the neutron regarded as
being static, i.e., $E_n = m_n +{\cal O}(\mN^{-1} )$.
To NLO, the above expression for $f(E)$ needs to be corrected to incorporate 
the kinetic energy of the recoil neutron from the relation 
$E_n=m_n \!+\!(\boldp_\nu\!-\!\boldp_e)^2/(2\mN)\! +\! {\cal O}(\mN^{-2})$. 
Corresponding to this change in $E_n$, we have
\begin{eqnarray} 
 E \!&=&\!\! E_\nu\! -\!\Delta_N\! -\! 
 (\boldp_\nu\!-\!\boldp_e)^2/(2m_n) +\cdots = 
\tilde{E}-(\boldp_\nu\!-\!\boldp_e)^2/(2m_n) +\cdots 
\nonumber \\ 
\!& = & \!\!
 \tilde{E} \left[ 1-\!\frac{1}{\mN}\left( E_\nu (1\!-\!\tilde{\beta}\cos\theta_e ) 
+\frac{\Delta_N^2-m_e^2}{2\tilde{E}} \right)  + {\cal O}(\mN^{-2}) \right] \; , 
\label{eq:recoil3}
\end{eqnarray} 
where, as earlier, 
$\tilde{E}=E_\nu -\Delta_N$, and the positron velocity becomes 
\begin{eqnarray}
\beta\!&=&\!\! \tilde{\beta}
\left[ 1-\!\frac{1}{\mN}\!\left(\frac{1\!-\!\tilde{\beta}^2}{\tilde{\beta}^2}\right)\!
\left( E_\nu (1\!-\!\tilde{\beta}\cos\theta_e ) +\frac{\Delta_N^2-m_e^2}{2\tilde{E}}\right) \!
+{\cal O}(\mN^{-2}) 
\right] \!, 
\label{eq:beta}
\end{eqnarray}
where $\tilde{\beta}\!=\!\sqrt{\tilde{E}^2-m_e^2}/\tilde{E}$. 
Note that the positron energy, $E$ and the velocity, $\beta$, are equal 
to the recoil-corrected $E_e^{(1)}$ and $v_e^{(1)}$ in Ref.~\cite{VB1999}, 
respectively. 
Reflecting these changes, 
the phase space integral in Eq.(\ref{eq:NLO}) 
needs to be corrected as follows:
\begin{eqnarray}
\int ({\rm d}{\cal F}) \ \frac{ f(E) }{4\pi} 
\left(\frac{m_n+E_n}{E_n}\right) 
\delta ( {\cal F} ) 
\left(\left| \frac{ {\rm d}{\cal F} }{{\rm d} E }\right|_{{\cal F}=0}\right)^{-1}  
\Big[ (1+3g_A^2) +(1-g^2_A )\beta \cos\theta_e \Big] \,,  
\label{eq:recoil2}
\end{eqnarray}
where ${\cal F}= E_\nu-\Delta_N -E -(\boldp_\nu -\boldp_e)^2/(2\mN) +\cdots$.
The factor $(m_n+E_n)/E_n \simeq 2$ in Eq.(\ref{eq:recoil2})   
has corrections of order $\mN^{-2}$, and the Jacobian factor 
%
produces the following  NLO phase space factor in Eq.(\ref{eq:NLO}):
\begin{eqnarray}
f(E) =  
\frac{E^2\beta}{\pi} \left[ 
1-\frac{E}{m_N}\left(1-\frac{E_\nu}{\beta E} \cos \theta_e\right) +
{\cal O}(m_N^{-2})  \right] \; ,
\label{eq:PS-recoil}
\end{eqnarray}
where the expressions for $E$ and $\beta$ are given in 
Eqs.~(\ref{eq:recoil3}) and (\ref{eq:beta}).

\vspace{7mm}
\noindent
{\bf Corrections from the next-to-leading-order lagrangian}\\

The $\mN^{-1}$ corrections to the Lagrangian
are explicitly written in Eq.(\ref{eq:Lnnee}). 
As noted before,  the radiative corrections 
to these additional terms in the Lagrangian
are of higher order than NLO in our counting and hence   
need not be considered here.
Evaluating the NLO lagrangian recoil correction contributions,  
illustrated in diagram (a) in Fig.~\ref{fig:feynmanQED}, we obtain the recoil
terms in Eq.(\ref{eq:Gi})
\begin{eqnarray}
{\cal G}_1^{recoil} (\beta)&=& \beta^2E\left(\frac{1-2\gA\muV+\gA^2}{1+3\gA^2}\right) 
- E_\nu \left(\frac{1+2\gA\muV+\gA^2}{1+3\gA^2}\right) 
\nonumber \\ 
{\cal G}_2^{recoil} (\beta)&=& E\left(\frac{1+2\gA\muV+\gA^2}{1-\gA^2}\right) 
- E_\nu \left(\frac{1-2\gA\muV+\gA^2}{1-\gA^2}\right) \; . 
\label{eq:L-recoil}
\end{eqnarray}
Comparing these results with those obtained 
for neutron $\beta$-decay~\cite{Ando2004},
we note that  there are several relative sign differences.\footnote{
This is in contrast to the order-$\alpha$ RCs 
which are universal at NLO in effective field theory. 
}
 Apart from the $m_N^{-1}$ phase-space corrections 
in neutron $\beta$-decay, 
the  $m_N^{-1}$ corrections 
(arising from the Lagrangian $m_N^{-1}$ interaction terms) 
relevant to the neutron life-time are 
contained in the $C_0$ factor in Eq.(14) of Ref.~\cite{Ando2004}. 
Noting  that in neutron beta-decay 
$E_e^{max}=E_\nu+E_e +{\cal O}(m_N^{-1})$,  
we may rewrite the $C_0$ factor as
\begin{eqnarray}
C_0(E_e) &=& 1+\frac{1}{m_N }\left\{ 
\beta^2 E_e \left(\frac{1+2\mu_Vg_A+g_A^2}{1+3g_A^2}\right) 
+E_\nu\left(\frac{1-2\mu_Vg_A+g_A^2}{1+3g_A^2}\right)
\right\}
\nonumber
\end{eqnarray}
Comparison of $C_0$ with ${\cal G}_1^{recoil} (\beta)$ 
in Eq.(\ref{eq:L-recoil}) 
clearly indicates that the $\mN^{-1}$ recoil corrections 
are not identical for the neutron $\beta$-decay and 
the inverse $\beta$-decay.
Moreover, since the weak-magnetism term is dominant,
the difference between $C_0$ and ${\cal G}_1^{recoil} (\beta)$ 
are of the same magnitude 
as the corrections themselves. 

Combining the $\mN^{-1}$ Jacobian factor
in the square brackets in Eq.~(\ref{eq:PS-recoil}), 
and the recoil correction arising from the lagrangian,
Eq.~(\ref{eq:L-recoil}),   
we confirm the recoil corrections given in 
Eqs.~(12) and (13) in Ref.~\cite{VB1999}.  
We prefer to keep these two $\mN^{-1}$ corrections separate since one is 
of a kinematical origin (phase space correction), 
whereas the other is of a dynamical origin arising from the 
transition matrix element.

\section{ Discussion} 

In this paper we have derived a model-independent 
expression for the 
radiative and 
$\mN^{-1}$  
corrections for the low-energy anti-neutrino proton reaction 
to next-to-leading-order 
in an effective field theory approach. 
We have shown that short-distance physics
not probed in this low-energy reaction can be subsumed into 
a single low-energy constant $\eV^R(\mu^2)$.   
In the $\bar{Q}/\Lambda_{\chi}$-expansion scheme adopted here,
the ${\cal O}(\alpha)$ and ${\cal O}(Q/m_N)$ corrections 
are considered to be of the same order 
for the reactor anti-neutrino  energy range.
We have found that the 
$\mN^{-1}$ corrections appearing in Eq.~(\ref{eq:L-recoil}), 
which originate from the lagrangian Eq.~(\ref{eq:Lnnee}), 
are different from the $\mN^{-1}$ corrections found in neutron 
$\beta$-decay, see e.g., Ref.~\cite{Ando2004}.
Therefore, to the order under consideration,
it is not advisable to write the 
inverse beta-decay 
cross section (or the positron yield) 
in terms of the neutron mean life $\tau_n$, 
as advocated in Ref.~\cite{Vogel84}.

The short-distance hadronic physics associated with the LEC, 
$\eV^R(\mu^2)$, 
was extensively discussed in Refs.~\cite{Sirlin1974,MS86}. 
The processes involved in $\eV^R(\mu^2)$ were studied from an effective 
field theoretical perspective in Ref.~\cite{Ando2004}. 
In principle, we should be able to determine the LEC, $\eV^R(\mu^2)$,
from high-precision experimental data.  
Relegating this determination to future study,
we choose here to estimate $\eV^R(\mu^2)$ at the scale $\mu$ = $\mN$
by comparing the short-distance radiative 
corrections calculated in the S-M approach~\cite{Sirlin1974,MS86} 
and the expressions obtained in EFT in Ref.~\cite{Ando2004}.
The result is 
\begin{eqnarray}
\tilde{e}_V^R(\mN^2) =
4\,{\rm ln}\!\left(\!\frac{m_Z}{m_p}\!\right)
\!+\!{\rm ln}\!\left(\!\frac{m_p}{m_A}\!\right)\!+\!2C \!+\!A_g =
18.31-.25+1.78-0.34=19.5  ,
\label{eq:LEC}
\end{eqnarray}
where, for the sake  of definiteness, 
the value of the axial matching mass $m_A$=$1.2$ GeV has been used
although its value involves uncertainty~\cite{Sirlin1974,MS86}. 
With this value of $\tilde{e}_V^R(\mN^2)$,
the correction term involving LEC in Eqs.~(\ref{eq:G1}) and (\ref{eq:G2}) 
is estimated to be $(\alpha /2\pi) \ \tilde{e}_V^R(\mN^2) \simeq 0.023$. 
The dominant first term in Eq.~(\ref{eq:LEC}) arises from well-known additional 
box diagrams with  Z-exchange, replacing the  photon-exchange, 
in electro-weak theory~\cite{Sirlin1974,MS86}.
This electro-weak physics can be naturally included in our approach. 
However, for an easy comparison with the neutron beta-decay radiative corrections 
evaluated in Ref.~\cite{Ando2004}, 
we prefer to keep this contribution in the above LEC. 
As for the last two terms in Eq.~(\ref{eq:LEC}), 
we remark that $A_g$ involves genuine short-distance hadron-structure physics,
whereas the constant $C$ arises from photon-loop diagrams in which 
the photon couples to the nucleon magnetic moments and also from     
the hadronic form factors. 
The long-range parts of these corrections are naturally included in EFT
at higher orders than considered in this paper. 

As a final comment we note that 
in our work we have used the value of the Fermi constant  $G_F$ 
determined from the muon lifetime measurement. 
The theoretical expression for $G_F$ is
evaluated in standard electroweak theory, and it
naturally includes log-terms involving $m_{Z}$.
These log-terms appear in our  expression for 
$\eV^R(\mu^2)$, Eq.~(\ref{eq:LEC}),
and was also considered in Ref.~\cite{Ando2004}, see e.g., 
Refs.~\cite{Sirlin1974,MS86} for details. 

In summary the integrated cross section for 
 reaction (\ref{eq:reaction}) 
is
\begin{eqnarray}
\sigma &=& (G_F V_{ud})^2 \frac{\tilde{E}^2\tilde{\beta}}{\pi} (1+3g_A^2)
\left(1+\frac{\alpha}{2\pi} {\cal G}_1^{rad} (\tilde{\beta})\right)  
\nonumber \\ 
&& \times \left\{ 1+\frac{1}{m_N}\Big[ {\cal G}_1^{recoil} (\tilde{\beta}) 
-\tilde{E} - \left(\frac{1+ \tilde{\beta}^2}{\tilde{\beta}^2} \right)
\left(E_\nu +\frac{\Delta_N^2-m_e^2}{2\tilde{E}}\right) 
\right.
\nonumber \\ && 
\left.  
+\left(\frac{1-g_A^2}{1+3g_A^2}\right)\frac{E_\nu}{3 }(2+\tilde{\beta}^2)
\Big] \right\}
\label{eq:sigma}
\end{eqnarray}
where as before, $\tilde{E} = E_\nu-(m_n-m_p)$ and 
$\tilde{\beta} = \sqrt{\tilde{E}-m_e^2}/\tilde{E}$, and where all $m_N^{-1}$ corrections 
in Eq.(\ref{eq:sigma}) except 
${\cal G}_1^{recoil} (\beta) $ of Eq.(\ref{eq:L-recoil}) originate from the 
phase-space factor $f(E)$, Eq.(\ref{eq:PS-recoil}). 

\vspace{5mm}

{\bf Acknowledgements}
We are grateful to V. Gudkov and T. Kubota for useful discussion. 
This work is  supported in part by the National Science Foundation grants,
PHYS-0758114 and PHY-1068305.

\section*{Appendix }

We use dimensional regularization to isolate the IR singularity. 
The 3-dimensional integral over $\tilde{\boldk} $ in Eq.(\ref{eq:sigmabr}) 
is replaced with a 
$d=4-2\epsilon $ dimensional integral where 
$\epsilon < 0$ for the purpose of handling the IR singularity, i.e.  
Eq.(\ref{eq:sigmabr}) is rewritten as 
\begin{eqnarray} 
\frac{ {\rm d}\sigma_{br}}{ {\rm d}{\rm cos}\theta_e}\!\!\! &=& \!\!\! 
\frac{ 1}{  32m_n m_p E_\nu }\  \frac{\mu^{4-d}}{(2\pi)^{d}} 
\int_0^{\tilde{E}-m_e} \!\!\!{\rm d}|\tilde{\boldk} | \  |\tilde{\boldk} |^{d-2} \!
{\rm d}^{d-2}\Omega_{\tilde{\boldk}} \!\left(\!\frac{ |\tilde{\boldp}_e | }{ |\tilde{\boldk} |}\!\right)  
\frac{1}{2}\sum_{spin} |{\cal M}_{br} |^2 \; , 
\label{eq:sigmaBR}
\end{eqnarray}
where $\tilde{\boldp}_e =\sqrt{(\tilde{E}\!-\!|\tilde{\boldk} |)^2\!-\!m_e^2}$. 
We note that in dimensional regularization,  the angular integration 
$ \int {\rm d}^{d-2}\Omega_{\tilde{\boldk}} $ yields 
($\hat{\tilde{\boldp}}_e\!\cdot\! \hat{\tilde{\boldk}}= \cos \theta_k$)
\begin{eqnarray}
\lefteqn{ \frac{\mu^{4-d}}{(2\pi)^{d}}  
\int \frac{ {\rm d}^{d-2}\Omega_{\tilde{\boldk}} [1- {\rm cos}^2\theta_k ] } 
{[1-\tilde{\beta} {\rm cos}(\theta_k )]^2}  }
\nonumber \\
 &=& \frac{\mu^{4-d}}{8\pi^3} 
\left\{  \Big[ 1\!+\! |\epsilon | \left(\gamma_E \!-\! {\rm ln}(4\pi) \right) \Big] 
\left[ -\frac{4}{\tilde{\beta}^2} +
\frac{2}{\tilde{\beta}^3}{\rm ln}\left(\frac{1\!+\!\tilde{\beta}}{1\!-\!\tilde{\beta}}\right)   \right] 
+ \frac{4 |\epsilon |}{\tilde{\beta}^2} \ {\cal C}(\tilde{\beta}) +{\cal O}(\epsilon^2)\right\} \; , 
\nonumber
\end{eqnarray} 
where $\tilde{\beta}\!=\!\sqrt{\tilde{E}^2-m_e^2}/\tilde{E}$ and the 
function ${\cal C}(\tilde{\beta})$ is given by (see e.g. Refs.~\cite{MS75}) 
\begin{eqnarray}
{\cal C}(\tilde{\beta})\!\!&=& \!\!1+\frac{1}{2\tilde{\beta}}
{\rm ln}\!\left(\frac{1\!+\!\tilde{\beta}}{1\!-\!\tilde{\beta}}\right) \!
\left[ 1-\frac{1}{2}\,{\rm ln}\!\left(\frac{1\!+\!\tilde{\beta}}{1\!-\!\tilde{\beta}}\right)\right] 
\nonumber\\
&&
\,\,+\,\,\,2\,{\rm ln}2
\left[ \frac{1}{2\tilde{\beta}}{\rm ln}\!\left(\frac{1+\tilde{\beta}}{1-\tilde{\beta}}\right) -1\right]
+ \frac{1}{\tilde{\beta}} L\!\left(\!\frac{2\tilde{\beta}}{1+\tilde{\beta}}\!\right)\, ,
\end{eqnarray}
and $L(x)$ is the Spence function 
\begin{eqnarray}
L(x)= - Li_2(x)= \int_0^x {\rm d}t \frac{{\rm ln}(1-t)}{t} \; . 
\nonumber 
\end{eqnarray} 
The integral over the photon momentum $\int {\rm d} \tilde{k}\ \tilde{k}^{d-5} \propto 1/|\epsilon |$ 
exhibit the IR-singularity. 
When we combine our expression for the 
integrated bremsstrahlung cross section with the contributions from the virtual 
photon loops, we find that the IR singularity is removed as it should.


\end{document}